\documentclass[prb,twocolumn,showpacs,amsmath,nobibnotes,nofootinbib,floatfix]{revtex4}
\usepackage[latin1]{inputenc}
\usepackage{color}
\usepackage{graphicx}
\usepackage{bm}
\usepackage{amssymb}
\usepackage{amsmath}
\newcommand{\Tr}{\mathop{\rm Tr}\nolimits}

\begin{document}
\title{Quantum and classical multiple scattering effects in spin dynamics of cavity polaritons}

\author{M.~M.~Glazov}
\email{glazov@coherent.ioffe.ru} \affiliation{A. F. Ioffe
Physico-Technical Institute, Russian Academy of Sciences, 194021
St.~Petersburg, Russia}
\author{L.~E.~Golub}
\affiliation{A. F. Ioffe
Physico-Technical Institute, Russian Academy of Sciences, 194021
St.~Petersburg, Russia}

\begin{abstract}
The transport properties of exciton-polaritons are studied with allowance for their polarization. Both classical multiple scattering effects and quantum effects such as weak localization are taken into account in the framework of a generalized kinetic equation. The longitudinal-transverse~(TE-TM) splitting of polariton states which plays role analogous to the spin-orbit splitting in electron systems is taken into account. The developed formalism is applied to calculate the particle and spin diffusion coefficients of exciton-polaritons, spin relaxation rates and the polarization conversion efficiency under the conditions of the optical spin Hall effect. In contrast to the electron systems, strong spin splitting does not lead to the antilocalization behavior of the particle diffusion coefficient, while quantum corrections to spin diffusion and polarization conversion can be both negative and positive depending on the spin splitting value.
\end{abstract}


\maketitle






\section{Introduction}\label{sec:intro}

Spin dynamics of charge carriers and their complexes attracts lately an increasing interest. The issues of spin coherence generation, detection and manipulation became topical during last years.~\cite{zutic:323}

The spin properties of cavity polaritons are of special interest both from the fundamental point of view and due to possible future device applications.~\cite{kavokin03b} The quantum microcavity is the quantum well embedded between two highly reflective Bragg mirrors. In such structures the strong coupling between a cavity photon and a quantum well exciton takes place, which leads to the formation of new quasi-particles: exciton-polaritons. These half-light half-matter particles exhibit both photonic and excitonic properties. Their polarization (or spin) dynamics is extensively studied both experimentally and theoretically, see Ref.~\onlinecite{Solnyshkov07} and references therein.

The polarization eigenmodes of the quantum microcavity are so-called TE- and TM-modes where the electric or magnetic field vector is oriented perpendicularly to the polariton wave vector, respectively. They are splitted by the longitudinal-transverse (also known as TE-TM) splitting~\cite{panzarini99} which plays a role similar to the spin splitting of electron states in quantum wells.~\cite{kavokin05a} It leads to D'yakonov-Perel'-like spin relaxation in the collision-dominated regime and to spin precession in the collision-free regime.~\cite{kavokin:017401,martin} One of the brightest manifestations of the polariton spin splitting is the polarization conversion or the optical spin Hall effect: under the Rayleigh scattering of linearly polarized polaritons the scattered particles obtain a certain degree of circular polarization.~\cite{kavokin05a,leyder07} The angular distribution of the circular polarization demonstrates the second angular harmonics thus reflecting the symmetry of the longitudinal-transverse splitting.

Spin splitting can strongly modulate the interference phenomena inherent to the quantum particles. It is well known that the spin-orbit  interaction modifies quantum corrections to electron diffusion coefficient or conductivity and electron spin relaxation times.~\cite{hikami80,aa,lyubinskiy04} Although the quantum interference of excitons has a long history,~\cite{ivch_pikus,PhysRevB.52.5233} it became topical only recently with the development of microcavities.~\cite{Savona00,gurioli:183901}

Here we analyse the spin-dependent interference effects in quantum microcavities. We focus on weak localization effects in exciton-polariton diffusion and spin diffusion, effects of interference in spin relaxation and in polarization conversion.
Our results can be summarized as follows:
\begin{enumerate}
 \item The quantum correction to the polariton diffusion coefficient is negative despite the value of the longitudinal-transverse splitting. It is in sharp contrast with the case of electrons, where the sufficiently strong spin-orbit interaction changes the sign of quantum correction to the diffusion coefficient.
 \item The relaxation of the circular polarization degree of exciton-polaritons is enhanced by the quantum interference effects while the relaxation of linear polarization can either speed up or slow down depending on the value of the longitudinal-transverse splitting.
 \item The efficiency of the polarization conversion can either be increased or decreased by the interference effects depending on the value of the longitudinal-transverse splitting and relation between the scattering time and the radiative lifetime of polaritons.
\end{enumerate}

The paper is organized as follows: in Sec.~\ref{sec:model} we present the model based on the kinetic equation for exciton-polariton spin density matrix. The quantum corrections to the collision integral describing the effects of coherent scattering are introduced and calculated in the framework of Green's function technique. Section~\ref{sec:diffusion} is devoted to the calculation of quantum corrections to the particle and spin diffusion coefficients of exciton-polaritons. The interference effects on exciton-polariton spin relaxation times are discussed in Sec.~\ref{sec:spinrel}. The multiple scattering effects and quantum interference effects in the polarization conversion are discussed in Sec.~\ref{sec:oshe}.

\section{Theory}\label{sec:model}

Below we present the kinetic theory of the spin dynamics of exciton-polaritons with allowance for the interference effects.

\subsection{Model}

We consider exciton-polaritons formed from the heavy-hole quantum well excitons. Their spin projection on the structure growth axis $z$ can take values $\pm 1$ or $\pm 2$. The latter states are optically inactive and do not participate to the light-matter coupling while the former ones constitute the radiative doublet. It is convenient to describe this doublet as a pseudospin $1/2$ state,~\cite{maialle93} the pseudospin $z$ component describes the emission intensity in the circular polarization, and the in-plane components correspond to the linear polarization: namely, $s_x$ component is proportional to the intensity measured in the given axes $x-y$ while $s_y$ component corresponds to the intensity measured in the axes $x'-y'$ rotated by $\pi/4$ with respect to $x-y$ coordinate frame.

The (pseudo)spin dynamics of exciton-polaritons is most conveniently described within the spin density matrix approach. It can be represented in a form
$$\rho_{\bm k} = f_{\bm k} + \bm s_{\bm k} \cdot \bm \sigma,$$
where $f_{\bm k}$ is the particle distribution function and $\bm s_{\bm k}$ is the average spin in the given state $\bm k$. Under the conditions of Rayleigh scattering experiments the monoenergetic distribution of the polaritons is excited and the processes of energy relaxation can be neglected.~\cite{kavokin03b} It means that the absolute values of polariton wave vectors $k_0$ are conserved. We consider the situation where the mean free path of the particles $l$ is large enough, $k_0 l\gg 1$. In this regime the dynamics of scattered particles can be described in the framework of the classical kinetic equation, and the quantum effects can be incorporated as corrections to the collision integral.~\cite{dmitriev97}

In the steady-state regime the kinetic equations writes:
\begin{eqnarray}
\frac{f_{\bm k}}{\tau_0} + Q\{f_{\bm k}\} = g_{\bm k}, \label{fk_g}\\
\frac{\bm s_{\bm k}}{\tau_0} + \bm s_{\bm k} \times \bm \Omega_{\bm k} + \bm Q\{\bm s_{\bm k}\} = \bm g_{\bm k}. \label{sk_g}
\end{eqnarray}
Here $\tau_0$ is the polariton lifetime, $Q\{f_{\bm k}\}$ and $\bm Q\{\bm s_{\bm k}\}$ are the collision integrals, and $g_{\bm k}$, $\bm g_{\bm k}$ are the components of the generation density matrix $\gamma_{\bm k} = (g_{\bm k} + \bm g_{\bm k}\cdot \bm \sigma)$ describing the particle and spin generation rates, respectively.

The quantity $\bm \Omega_{\bm k}$ in Eq.~\eqref{sk_g} is the pseudo-spin precession frequency related to the longitudinal-transverse splitting of polariton modes equal to $\hbar |\bm \Omega_{\bm k}|$.~\cite{panzarini99} It can be written as
\begin{equation}\label{omega_k}
\bm \Omega_{\bm k} = \Omega(k) [\cos{2\varphi_{\bm k}}, \sin{2\varphi_{\bm k}},0],
\end{equation}
where $\Omega(k)$ is some function of the wave vector absolute value $k$, $\varphi_{\bm k}$ is the angular coordinate of $\bm k$. It depends strongly on the microcavity parameters.  Note, that the components of $\bm \Omega_{\bm k}$ are described by the \emph{second} angular harmonics of the wave vector angle, because the pseudospin flip is accompanied with the change of the polariton spin by two. It is strongly different from the case of two-dimensional electrons, where the spin splitting is described by the \emph{first} and \emph{third} angular harmonics.

We assume that the scattering of exciton-polaritons is caused by a short-range disorder, i.e. the scattering cross-section is angular independent. This condition can be violated in real structures,~\cite{savona07} but our goal is to consider the simplest case which allows an analytical solution. Furthermore, we assume that the polaritons are described by a parabolic dispersion, $E_{\bm k} = \hbar^2k^2/2m$, with an effective mass $m$, and introduce the density of states per spin $\mathcal D = m/2\pi\hbar^2$.

In typical microcavities under the conditions of Rayleigh scattering the kinetic energy of polaritons, $E_{\bm k}$, is of the order of several meV, while the longitudinal-transverse splitting is of the order of tenths of meV. Therefore the effect of the polarization on the orbital dynamics of polaritons can be neglected. At the same time, the spin precession frequency, inverse lifetime and the scattering rate can be comparable.

The collision integrals entering Eqs. \eqref{fk_g}, \eqref{sk_g} can be written as a sum of the classical contribution
\begin{eqnarray}\label{q_cl}
Q_{cl}\{f_{\bm k}\} &=& Q\sum_{\bm k'} (f_{\bm k} - f_{\bm k'}) \delta (E_{\bm k} -E_{\bm k'}), \\
\bm Q_{cl}\{\bm s_{\bm k}\} &=& Q\sum_{\bm k'} (\bm s_{\bm k} - \bm s_{\bm k'}) \delta (E_{\bm k} -E_{\bm k'}), \nonumber
\end{eqnarray}
and the quantum corrections. Here $Q$ is the elastic scattering constant, $Q\mathcal D = \tau_1^{-1}$ with $\tau_1$ being the momentum scattering time.

\begin{figure}
\centering
\includegraphics[width=\linewidth]{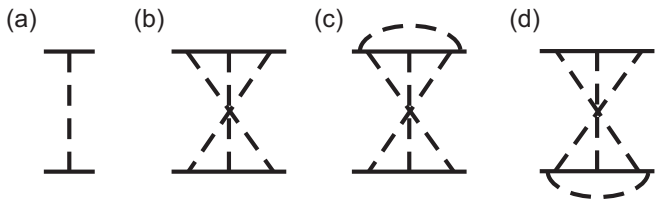}
\caption{Examples of irreducible diagrams which contribute to the collision integral. (a) Single scattering, (b) coherent backscattering, (c) and (d) coherent scattering by an arbitrary angle.}\label{fig:diag}
\end{figure}

The quantum corrections to the collision integrals can be most conveniently found in Green's function technique. Various contributions to the scattering cross-sections are exemplified in Fig.~\ref{fig:diag}.
Solid lines are retarded and advanced exciton-polariton Green's functions which, with allowance for polariton spin and the longitudinal-transverse splitting have a form of $2\times 2$ matrices and read~\cite{iordanskii94}
\begin{equation}\label{greens}
\mathcal G^{R,A}(\bm k, \omega) = [\hbar \omega - E_{\bm k} - {\hbar}(\bm \sigma \cdot \bm \Omega_{\bm k})/2 \pm {\mathrm i\hbar}/{2\tau}]^{-1},
\end{equation}
where
\[
\tau^{-1} = \tau_0^{-1} + \tau_1^{-1}.
\]
Note that the radiative lifetime of exciton-polaritons plays role of the phase relaxation time in the theory of electron weak localization.~\cite{aa}
The dashed line in Fig.~\ref{fig:diag} is the scattering amplitude which
reads $\hbar Q/2\pi = \hbar^3/m\tau_1$.

The diagram Fig.~\ref{fig:diag}(a) shows the single scattering process. The diagram in Fig.~\ref{fig:diag}(b) describes the interference of polariton which passes the same configuration of scatterers in the clockwise and counterclockwise directions and propagates exactly backwards. Diagrams Fig~\ref{fig:diag}(c) and (d) describe the same interference but accompanied by the scattering by an arbitrary angle. The corrections to the collision integral can be expressed in the terms of the Cooperon operator $\mathcal C^{\alpha\beta}_{\delta\gamma} (\bm q)$ which is the sum of diagrams depicted in Fig.~\ref{fig:diag}(b) with any number of lines $N\geqslant 3$.(\footnote{The diagram with two intersecting impurity lines is dominated by the pairs of impurities separated by $\sim k_0^{-1}$ and can not be treated within our approximation $k_0 l\gg 1$.}) The Cooperon satisfies the equation
\begin{equation}\label{cooperon}
{\cal C}^{\alpha\beta}_{\delta\gamma}(\bm q) =
[\mathcal P^3(\bm q)]^{\alpha\beta}_{\delta\gamma} + \sum_{\beta'\gamma'}
    {\cal C}^{\alpha\beta'}_{\delta\gamma'}(\bm q) \mathcal P^{\beta'\beta}_{\gamma'\gamma}(\bm q),
\end{equation}
with
\begin{equation}\label{p}
\mathcal P^{\alpha\beta}_{\delta\gamma}(\bm q) = \frac{\hbar Q}{2\pi} \sum_{\bm k'} \mathcal G^R_{\alpha\beta}(\bm k',E_0)\mathcal G^A_{\delta\gamma}(\bm q - \bm k',E_{0}).
\end{equation}
Here it is assumed that the polaritons have the same energy $E_0$ determined by the excitation. As a result the
distribution functions can be reduced to the angular dependent parts only
\begin{eqnarray}
f_{\bm k} = f(\varphi_{\bm k}) \delta(E_{\bm k} - E_{0})\nonumber,\\
\bm s_{\bm k} = \bm s(\varphi_{\bm k}) \delta(E_{\bm k} - E_{0})\nonumber.
\end{eqnarray}

The
shape of the coherent backscattering cone obtained by the summation of the diagrams in Fig.~\ref{fig:diag}(b) is given in the limit of $\Omega\tau =0$ by the following expression~\cite{ivch_pikus}
\begin{equation}\label{backscattering}
I^1(\bm k) = I^0(\bm k)\frac{\tau}{\tau_0}\frac{1}{\sqrt{(\tau_1/\tau)^2 + (\bm k + \bm k_0)^2l^2}-1}.
\end{equation}
Here $I^0(\bm k)$ describes the classical intensity distribution. In the case of exact backscattering $\bm k = - \bm k_0$ and $\tau_0 \gg \tau_1$ the coherent backscattering intensity is $I^1(\bm k) = I^0(\bm k)$. If $\tau_0 \ll \tau_1$ the coherent backscattering is negligible $I^1(\bm k) \ll I^0(\bm k)$.
The backscattering cone angular width is small, of the order of $(k_0 l)^{-1} \ll 1$ and the details of its shape are irrelevant for our consideration. Thus, the coherent backscattering processes simply couple states with $\bm k$ and $-\bm k$.

The quantum effects are most pronounced in the multiple scattering regime where $\tau_0 \gg \tau$. This assumption is used hereafter in the description of the interference phenomena.
Furthermore, in the limit $\ln(\tau_0/\tau_1) \gg 1$ the sum of the diagrams Fig.~\ref{fig:diag}(c) and (d) weakly depends on the scattering angle. Even if this condition is violated the classical effects of finite correlation length of the potential leading to the angular dependent scattering cross-section may be more important. Therefore their angular dependence is neglected, and the quantum corrections to the collision integral can be written as~\cite{dmitriev97,lyubinskiy04}
\begin{eqnarray}
Q_{qnt}\{f_{\bm k}\} = -
W_0 \int \frac{\mathrm d \varphi'}{2\pi} [f(\varphi - \pi) - f(\varphi')]\delta(E_{\bm k} - E_{0}), \label{delta_St_f}\\
\bm Q_{qnt}\{\bm s_{\bm k}\}= -
\hat W \int \frac{\mathrm d \varphi'}{2\pi} [\bm s(\varphi - \pi) - \bm s(\varphi')]\delta(E_{\bm k} - E_{0}). \nonumber
\end{eqnarray}
Here the quantities $W_0$, $\hat W$ describe spin-dependent return probabilities~\cite{aa}
\begin{equation}\label{hat_W}
W_0 = \frac{l}{k_0 \tau}\sum_{\alpha\beta} \sum_{\bm q} \mathcal C^{\alpha\beta}_{\beta\alpha}(\bm q),
\end{equation}
\[
(\hat W)_{ij} = \frac{l}{k_0\tau}\sum_{\alpha\beta\gamma\delta} \sum_{\bm q} \sigma^i_{\gamma\alpha}\mathcal C^{\alpha\beta}_{\delta\gamma}(\bm q)\sigma^j_{\beta\delta},
\]
where $\sigma^i$ for $i=x,y,z$ are the Pauli matrices.

We consider an isotropic spin splitting, i.e. $\Omega_{\bm k}$ is independent of the angles of the wave vector $\bm k$ and equals to $\Omega(k)$, see Eq. (\ref{omega_k}). Thus, for the cylindrical symmetry of the problem under study, the only non-zero components of $\hat W$ are
\[
 W_\perp = W_{xx} = W_{yy} , \quad W_\parallel = W_{zz}.
\]
Thus, the decription of the spin dynamics of the exciton-polaritons is reduced to the solution of the kinetic equations with the collision terms in the form of Eqs. \eqref{q_cl}, \eqref{delta_St_f}. Latter depend on the spin-dependent return probabilities $W_0$, $W_\parallel$ and $W_\perp$ which can be found straightforwardly from the Cooperon operator $\mathcal C^{\alpha\beta}_{\delta\gamma}(\bm q)$~Eq. \eqref{cooperon}.

\subsection{Cooperon}

In order to find the Cooperon we follow the procedure outlined in Refs.~\onlinecite{golub_wl_05,glazov_wl_06} and make use of the fact that the operator ${\cal P}$ can be presented as
\begin{equation}\label{aux_p_eq}
    {\cal P} = \frac{\tau}{\tau_1} \int\limits_0^{2\pi} \frac{\mathrm d \varphi_{\bm k}}{2 \pi}
    \left[ 1- {\rm i} \tau {\bm L} \cdot {\bm \Omega}_{\bm k} + {\rm i} \tau {\bm v}_{\bm k} \cdot {\bm q} \right]^{-1},
\end{equation}
where
$$\bm{L}_{\alpha\gamma,\beta\delta} =\frac{\bm{\sigma}_{\alpha\beta} - \bm{\sigma}_{\delta\gamma}}{2}$$
is an
operator of the difference of spins of two interfering particles and $\bm v_{\bm k}$ is the velocity operator. This result can be contrasted with the situation realized for two-dimensional electrons where the spin splitting is \emph{odd} function of $\bm k$ and, thus, the total spin of interfering particles enters into the definition of $\mathcal P$.

Nevertheless, in our treatment we use the representation of the total spin of the interfering particles:
$\alpha \gamma \to S m_s$,
where $S=0,1$ is the absolute value of the total spin $\bm S$,
and $m_s$ is its projection onto the $z$ axis ($|m_s| \leq S$).
The pair of particles with $S=0$ is in the singlet state while
$S=1$ corresponds to the triplet one.

Since ${\cal P}$ and, hence, the Cooperon are determined by the operator $\bm L$, two independent contributions to Cooperon can be separated. Namely, the matrix $\mathcal P$ can be block-diagonalized, one of the blocks corresponds to the pair in the triplet state with $m_s=0$ and another is a $3\times 3$ matrix corresponding to two triplet states with $|m_s|=1$ and a to singlet. The part with $(S,m_s) = (1,0)$ reads
\begin{equation}\label{P}
    P = \frac{1}{\sqrt{(\tau_1/\tau)^2 + (ql)^2}},
\end{equation}
and the corresponding Cooperon is given by
\begin{equation}\label{C0}
    {C}_0 = \frac{P^3}{1- P}.
\end{equation}

The operator ${\cal P}$ in the basis of three other states, $(S,m_s)= (1,1); (0,0); (1,-1)$, has the following form
\begin{equation}
\label{P1}
    {\cal P}_1 = \left(%
\begin{array}{ccc}
  P - S_{0} & {\rm i} e^{-2{\rm i} \varphi_{\bm q}} R & e^{-4{\rm i} \varphi_{\bm q}}S_4 \\
  {\rm i} e^{2{\rm i} \varphi_{\bm q}} R & P - 2S_{0} & -{\rm i} e^{-2{\rm i} \varphi_{\bm q}} R  \\
  e^{4{\rm i} \varphi_{\bm q}}S_4 & -{\rm i} e^{2{\rm i} \varphi_{\bm q}} R& P - S_{0} \\
\end{array}%
\right),
\end{equation}
where $\varphi_{\bm q}$ is the angular coordinate of $\bm q$,
\begin{eqnarray}\label{S}
    S_m &=& \int\limits_0^\infty dx \exp{(-x)} J_m \left(x ql \right) \sin^2{\left(\frac{x \Omega  \tau}{2}  \right)},\\
    R &=& \frac{1}{\sqrt{2}} \int\limits_0^\infty dx \exp{(-x)} J_2\left(x ql \right) \sin{(x\Omega\tau)}, \nonumber
\end{eqnarray}
and $J_m(y)$ are the Bessel functions.
(\footnote{The integrals~(\ref{S}) can be calculated analytically but the expressions are very cumbersome, therefore we leave them in integral forms.})
The Cooperon corresponding to these three states is given by
\begin{equation}\label{C1}
    {\cal C}_1 = {\cal P}_1^3 \left[ I - {\cal P}_1\right]^{-1},
\end{equation}
where $I$ is the $3 \times 3$ unit matrix.

\subsection{Spin-dependent return probabilities}

Using Eqs. \eqref{hat_W} and rewriting the expressions for $W_0$ and $W_{\|,\perp}$ in the same basis as $C_0$, $\mathcal C_1$, we have
\begin{eqnarray}
    W_0 &=& \frac{l}{k_0\tau} \sum_{\bm q} \left\{ \Tr{\left[\mathcal E_1 \mathcal C_1 \right]} + C_0 \right\}, \label{W0} \\
  W_\parallel &=& \frac{l}{k_0\tau} \sum_{\bm q} \left\{ \Tr{\left[\mathcal C_1 \right]} - C_0 \right\}, \nonumber \\
W_\perp &=& \frac{l}{k_0\tau} \sum_{\bm q} \left\{ \Tr{\left[\mathcal E_2 \mathcal C_1 \right]} + C_0 \right\}, \nonumber
\end{eqnarray}
where $\mathcal E_1$ is the matrix with the diagonal $(1,-1,1)$ and other elements being zero, and $\mathcal E_2$ is the matrix with unit anti-diagonal elements and other elements being zero. Note that $W_0$ can be recast in a conventional form as a difference of triplet and singlet contributions to $\mathcal C$.~\cite{aa,iordanskii94,golub_wl_05,glazov_wl_06}

\begin{figure*}[htb]
\centering
\includegraphics[width=\linewidth]{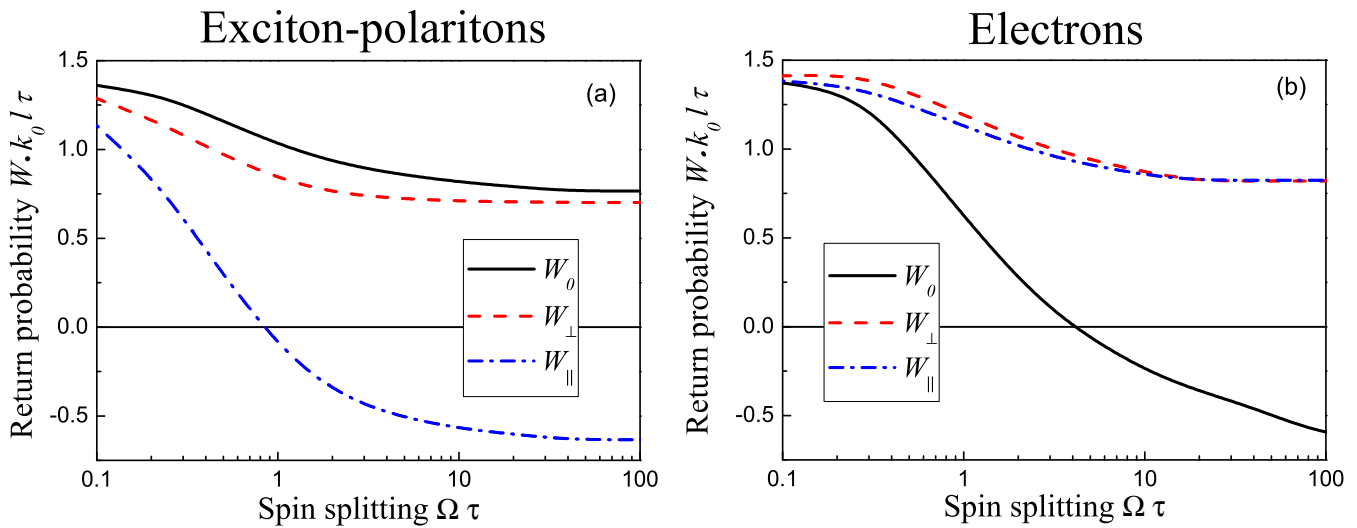}
\caption{(Color online) Components of the tensor $\hat W$ plotted as the functions of the product $\Omega\tau$, where $\Omega=\Omega(k_0)$. The radiative lifetime $\tau_0=100\tau_1$.}\label{fig:w}
\end{figure*}

The asymptotic values for the components of $\hat W$ tensor can be obtained analytically in the case $\ln{(\tau_0/\tau_1)} \gg 1$, $\Omega\tau_1\ll 1$. They read:
\begin{eqnarray}
\label{W0_diff}
W_0 &=& \frac{1}{2\pi \tau k_0 l} \left( 2 \ln{\frac{T_{s\perp}}{\tau}} + \ln{\frac{\tau_0}{\tau}} - \ln{\frac{T_{s\|}}{\tau}} \right),\\
W_\parallel &=& \frac{1}{2\pi \tau k_0 l}\left( 2 \ln{\frac{T_{s\perp}}{\tau}} + \ln{\frac{T_{s\|}}{\tau}} - \ln{\frac{\tau_0}{\tau}} \right),\nonumber \\
W_\perp &=& \frac{1}{2\pi \tau k_0 l} \left( \ln{\frac{T_{s\|}}{\tau}} + \ln{\frac{\tau_0}{\tau}} \right).\nonumber
\end{eqnarray}
Here the lifetimes are introduced for spin components parallel and perpendicular to the growth axis $z$:
\begin{equation}
    \frac{1}{T_{s\|}} = \Omega^2\tau + \frac{1 }{\tau_0}, \qquad \frac{1}{T_{s\perp}} = \frac{\Omega^2\tau}{ 2} + \frac{1}{\tau_0}.
\end{equation}

In this limit, the spin-dependent return probabilities contain logarithmic factors, which is a specific feature of two-dimensional systems where the return probability is proportional to the logarithm of the ratio of longest and shortest allowed travel times. The lower boundary for the travel time is obviously $\sim \tau$ while the upper boundary depends on the particle lifetime $\tau_0$ and spin lifetimes $T_{s\parallel}$, $T_{s\perp}$. The expression for the particle return probability $W_0$ differs from the corresponding transport coefficient derived for electrons in Ref.~\onlinecite{iordanskii94}. It reflects the specifics of the \emph{even} in $\bm k$ spin splitting in polariton systems.

In the regime of $\Omega\tau\gg 1$ the leading logarithmic contributions to $\hat W$ have the following form:
\begin{eqnarray}
W_0 = W_\perp \approx \frac{1}{2\pi\tau k_0 l}\left(0.06 + \ln{\frac{\tau_0}{\tau}}\right),\\
W_\parallel \approx \frac{1}{2\pi\tau k_0 l}\left(0.06 - \ln{\frac{\tau_0}{\tau}}\right) \nonumber.
\end{eqnarray}
Clearly, the main (logarithmically large) correction is determined by the $(S, m_s)=(1,0)$ contribution to the Cooperon. Other states give small additional term. It is worth noting that at $\Omega\tau\to\infty$ quantitites $W_0, W_\perp$ have the same sign as at $\Omega\to 0$, while $W_\parallel$ changes its sign as compared with the case $\Omega\tau\ll 1$.

The dependence of $W_0$, $W_\parallel$ and $W_\perp$ on spin splitting value in a wide range of $\Omega\tau$ variation is presented in Fig.~\ref{fig:w} at a fixed ratio $\tau_0/\tau_1=100$. Panel (a) of Fig.~\ref{fig:w} shows the non-zero components of $\hat W$ in the case of exciton-polaritons (i.e. where the spin splitting contains \emph{second} angular harmonics). The case of electrons where the spin splitting contains \emph{first} harmonics is presented for comparison in Fig.~\ref{fig:w}(b). Although the overall behavior of $W_\parallel$, $W_\perp$, $W_0$ is similar in both cases, there is a strong qualitative difference. Namely, for electrons $W_0$ changes its sign with an increase of $\Omega\tau$ which is a direct consequence of the antilocalization phenomena: the sufficiently strong spin-orbit interaction leads to the phase $\pi$ acquired by an electron on the closed trajectories, and the probability for electron to avoid its initial point increases.~\cite{glazov_wl_06} The real spin of polaritons is $1$ therefore the aquired phase is $2\pi$, and the quantity $W_0$ does not change its sign as a function of $\Omega\tau$. Mathematically, it is a direct consequence of the $\bm k$-even spin splitting: the interference is governed by the difference spin $\bm L$ of the particles contrary to the case of odd spin splitting where the interference is controlled by the total spin of the particles $\bm S$. Therefore, in the case of electrons the Cooperon can be separated in the singlet and triplet parts with respect to the total spin $\bm S$ which enter to $W_0$ with different signs.~\cite{hikami80} On the contrary, in the case of polaritons the part corresponding to the triplet with zero projection of the total spin $(S,m_s)=(1,0)$ is separated from the Cooperon, therefore $W_0$ keeps its sign, while $W_{\parallel}$ demonstrates the antilocalization behavior. Therefore, the antilocalization of polaritons \emph{does not occur}.


\begin{figure}[htb]
\centering
\includegraphics[width=\linewidth]{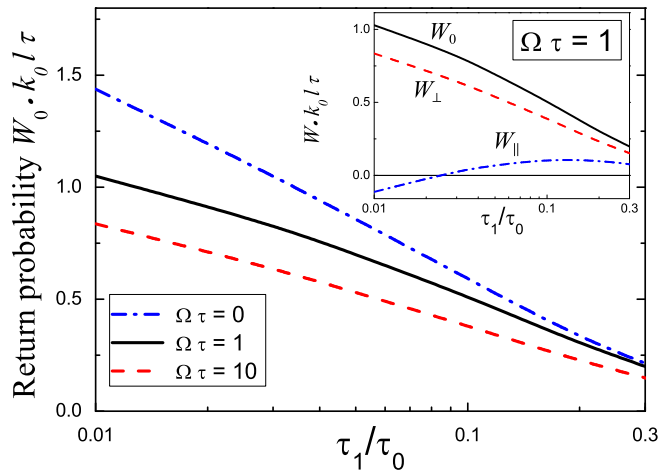}
\caption{(Color online) The dependence of $W_0$ on the ratio $\tau_1/\tau_0$ calculated for the different values of $\Omega\tau$. The inset shows the dependence of all components of $\hat W$ tensor at fixed $\Omega\tau=1$.}\label{fig:dephase}
\end{figure}

Figure~\ref{fig:dephase} shows the dependence of $W_0$ on the ratio $\tau_1/\tau_0$ calculated for exciton-polaritons. The curves correspond to the different values of $\Omega\tau$. It is seen that $W_0$ decreases monotonously with an increase of the radiative rate. An inset to the figure shows $W_0$, $W_\parallel$ and $W_\perp$ calculated as functions of $\tau_1/\tau_0$ at fixed $\Omega\tau=1$. In agreement with Eqs.~(\ref{W0})~and~(\ref{W0_diff}) only $W_\parallel$ behaves non-monotonously with the decrease of the polariton lifetime. Its dependence on $\tau_1/\tau_0$ is determined by the competition of two contributions to Cooperon: $\Omega$-independent $\mathcal C_0$ and $\Omega$-dependent $\mathcal C_1$. They enter into Eq.~(\ref{W0}) with the opposite signs and lead to non-monotonic behavior. Cooperons $\mathcal C_0$ and $\mathcal C_1$ contribute to $W_0$ and $W_\perp$ with the same signs and therefore these components of the tensor $\hat W$ are monotonous functions of $\tau_1/\tau_0$.

\section{Quantum corrections to the particle and spin diffusion coefficients}\label{sec:diffusion}

The diffusion coefficient $D_n$ for the quantity $n(\bm r, t)$
where $n$ can be either the particle density or one of the spin
density components $s_x, s_y$ and $s_z$ can be
defined from the Fick's law
\begin{equation}\label{Fourier}
D_n \nabla n + \bm j^n(\bm r, t) =0,
\end{equation}
where $\bm j^n$ is the flux density of the corresponding quantity, $\bm r$ is the coordinate, $t$ is time.

Let us first consider the particle diffusion coefficient $D_0$. We assume
that the distribution function of polaritons can be recast as
$f(\bm r, \bm k, t) = n(\bm r, t) f(\varphi_{\bm k}) \delta(E_{\bm
k} - \hbar \omega)$ and that $n(\bm r,t)$ weakly depends on
the coordinate and on time. The kinetic equation with the
collision integrals \eqref{q_cl}, \eqref{delta_St_f} reads
\begin{multline}
\frac{\partial n}{\partial t} f(\varphi_{\bm k})  +
 \frac{n}{\tau_0} f(\varphi_{\bm k}) + \frac{\partial n}{\partial
\bm r} \cdot \frac{\hbar \bm k}{m} f(\varphi_{\bm k}) +
\frac{n}{\tau_1} \Bigl[f(\varphi_{\bm k}) - \bar f\Bigr]\\ - n W_0
\Bigl[f(\varphi_{\bm k} - \pi) - \bar f\Bigr]=0.
\end{multline}
Here $\bar f = (2\pi)^{-1}\int_0^{2\pi} f(\varphi) \mathrm d\varphi$ is the average number of particles. Obviously $f(\varphi)$ contains only zeroth and first harmonics in $\varphi_{\bm k}$. The particle current thus reads
\begin{equation}\label{current}
\bm j^n(\bm r, t) =  n(\bm r, t) \int\limits_0^{2\pi}
\frac{\mathrm d \varphi_{\bm k}}{2\pi} \frac{\hbar \bm k}{m}
f(\varphi_{\bm k}),
\end{equation}
and Eq. \eqref{current} together with Eq. \eqref{Fourier}
yields for the particle diffusion coefficient
\begin{equation}\label{D0}
D_0 = D_{cl} (1 - W_0 \tau_1),
\end{equation}
where $D_{cl} = \hbar^2k^2\tau_1/2m^2$ is the classical diffusion coefficient.

Analogously one can derive the diffusion coefficients for the spin components $s_z$ ($\parallel$) and $s_{x,y}$ ($\perp$):
\begin{equation}
D_{\parallel,\perp} = D_{cl}(1 - W_{\parallel, \perp} \tau_1).\label{Dz}
\end{equation}

From Eqs. \eqref{D0}, \eqref{Dz} one can see that the quantum corrections to the particle and spin diffusion coefficients are determined by the respective components of the tensor $\hat W$.
It follows from the previous section that $W_0$ is positive for
all values of $\Omega\tau$ and $\tau_1/\tau_0$. Therefore the
backscattering is enhanced and the quantum interference leads to
the decrease of the polariton diffusion coefficient,
cf.\eqref{D0}. The same holds for the in-plane spin diffusion
coefficient, $D_\perp$, which is determined by $W_\perp>0$. The
quantum correction to the longitudinal spin diffusion coefficient,
$D_\parallel$, is determined by $W_\parallel$ which can be either
positive or negative depending on $\Omega\tau$ and $\tau_1/\tau$.
In particular, $W_\parallel$ is negative for large $\Omega\tau$
and $\tau_0/\tau_1$, and quantum effects will lead to an increase
of the longitudinal spin diffusion coefficient. For small values
of $\Omega\tau$ longitudinal diffusion coefficient is decreased by
the quantum interference.

Quantum corrections to diffusion coefficients can be extracted
from the dependence of these coefficients on external
perturbations which introduce an extra dephasing of the particles.~\cite{aa} These effects can be incorporated into the effective lifetime $\tau_0$. For instance, the temperature variation modifies the rates of inelastic processes.  Since exciton is a neutral particle, the magnetic
flux through the trajectory is proportional to the electron-hole
separation. Hence the magnetic field affects the interference at
$l_B \sim a_{\mathrm B}$ where $l_B$ is the magnetic length and $a_{\mathrm B}$ is
Bohr radius.~\cite{arseev98} However, such a field is not classically weak, therefore strong diamagnetic effects are dominant, therefore the weak localization corrections to magnetodiffusion can be hardy separated.

In microcavities $\tau_0$ can be efficiently varied with incidence angle of
light proportionally to the photonic fraction in polariton. It is seen for Fig.~\ref{fig:dephase} that the presence of the
spin-splitting does not lead to the non-monotonous dependence of
$W_0$ (and, therefore, of the quantum correction to the diffusion
constant) on the lifetime. On the contrary, the antilocalization behaviour can be observed in $D_\parallel$

\section{Quantum corrections to the spin relaxation rates}\label{sec:spinrel}

The spin dynamics is known to be non-exponential with allowance for the quantum interference effects: weak localization leads to the appearance of the long-living tails in spin polarization.~\cite{lyubinskiy04,PhysRevB.52.5233}
It is convenient to determine the tensor of spin relaxation rates $\hat \Gamma$ from the balance equation for the total spin in the system at the steady-state spin pumping
\begin{equation}
 \label{balance}
\left(\tau_0^{-1} + \hat \Gamma \right)\bar{\bm s} = \bm g.
\end{equation}
Here $\bar{\bm s} = (2\pi)^{-1} \int_0^{2\pi} \bm s(\varphi)
\mathrm d \varphi$ is the angular average of the spin
distribution, and the spin generation is assumed to be isotropic
\[
\bm g_{\bm k} = \bm g\delta(E_{\bm k}-E_0).
\]

We represent the distribution functions separating terms of zeroth and first order in quantum corrections:
\begin{equation}\label{decomposition}
f(\varphi) = f^0(\varphi) + f^1(\varphi),\quad \bm s(\varphi) = \bm s^0(\varphi) + \bm s^1(\varphi).
\end{equation}
Here the upper index refers to the order in $W_0$, $\hat W$.

In order to solve the kinetic equations~\eqref{fk_g}, \eqref{sk_g} we note that the solution of the following equation
\begin{equation}\label{s_eq_simple}
\frac{\bm s(\varphi)}{\tau} + \bm s(\varphi) \times \bm \Omega(\varphi) = \bm F(\varphi),
\end{equation}
where $\bm \Omega(\varphi) = \bm \Omega_{\bm k}$ writes
\begin{multline}\label{s_phi1}
\bm s (\varphi) = \frac{\tau}{1 + \Omega^2\tau^2}\times \\ \left[\bm F(\varphi) + \tau \bm \Omega(\varphi) \times \bm F(\varphi) + \tau^2 \bm \Omega(\varphi)(\bm \Omega(\varphi) \cdot \bm F(\varphi)) \right].
\end{multline}

Neglecting the weak-localization effects (i.e. putting $W_0,
W_\parallel, W_\perp =0$) one arrives to Eq.~\eqref{s_phi1} for the spin
distribution function $\overline{\bm s^0}$ with
\begin{equation}\label{F_steady_homog}
\bm F = \frac{\overline{\bm s^0}}{\tau_1} + \bm g.
\end{equation}

It is enough to consider the cases where $\bm g$ is directed along the growth axis of the sample, $\bm g \parallel z$, or lies in the plane of the structure, because tensor $\hat \Gamma$ for cylindrical symmetry has only two independent components,
\[
 \Gamma_{\parallel} = \Gamma_{zz}, \quad \Gamma_{\perp} = \Gamma_{xx}=\Gamma_{yy},
\]
describing the longitudinal and transverse relaxation rates, respectively.

If $\bm g \parallel z$ the self-consistency equation for $\bar{\bm
s}$ gives $\overline{s_z^0} = {g_z} T_{s\|}$, and hence
\begin{equation}\label{s0z_homog}
\bm s^0(\varphi) =  \left[\bm g + \tau \bm \Omega(\varphi) \times
\bm g\right] T_{s\|}.
\end{equation}
Therefore, the quantum correction for $z$-component of spin reads
\begin{equation}\label{s1z_homog}
\overline{s_z^1} = - g_z T_{s\|}^2 (\Omega\tau)^2 W_\perp,
\end{equation}
and according to Eq.~\eqref{balance} the longitudinal spin relaxation rate is given by
\begin{equation}\label{tauzz}
\Gamma_{\parallel} = \Omega^2\tau \left(1 + \tau W_\perp\right).
\end{equation}
Since $W_\perp$ is positive the $z$-component spin relaxation is enhanced by the quantum interference effects. This equation shows the correction to $\tau_1$ in the longitudinal spin relaxation rate has an inverse sign as compared to the correction to the spin diffusion coefficient Eqs. \eqref{Dz}. It is due to the fact that the spin relaxation rate is governed by the relaxation of the second harmonic of the spin distribution function~\cite{maialle93} whereas the spin diffusion is determined by the relaxation of the first harmonic.

The calculation of the transverse relaxation rate yields
\begin{equation}\label{tau_perp}
\Gamma_\perp = \frac{\Omega^2\tau/2}{1+(\Omega\tau)^2/2} \left[1 + \frac{\tau W_\parallel - \tau W_\perp (\Omega\tau)^2/2}{1+(\Omega\tau)^2/2}\right].
\end{equation}
This equation shows that at small $\Omega\tau$ the quantum corrections to $\Gamma_\perp$ are determined by $W_\parallel>0$ and the spin relaxation is enhanced by an interference. In contrast, at large $\Omega\tau_1$ the quantity $W_\parallel$ becomes negative and slows dows spin relaxation. Besides, the second contribution proportional to $W_\perp(\Omega\tau_1)^2>0$ becomes even more important and suppresses spin relaxation as well.

\section{Optical Spin Hall Effect}\label{sec:oshe}

Under the conditions of the optical spin Hall effect the TE- or
TM-eigenstate of the quantum microcavity with the wave vector $\bm
k_0$ is excited and the circular polarization in the scattered
state $\bm k$ is observed.~\cite{kavokin05a} In this case the generation
rate can be represented as
\begin{equation}\label{generation}
g_{\bm k}=g\delta_{\bm k, \bm k_0}, \quad \bm g_{\bm k}=\bm g
\delta_{\bm k, \bm k_0},
\end{equation}
where the Kronecker $\delta$-symbol used here is defined as
\[
\delta_{\bm k, \bm k_0} = \frac{2\pi}{\mathcal D} \delta(E_{\bm k} - E_{\bm k_0}) \delta(\varphi_{\bm k} - \varphi_{\bm k_0}).
\]

We first consider \emph{classical} multiple scattering effects in the optical spin Hall effect regime and demonstrate that the spin relaxation decreases the polarization degree as compared with the single scattering regime considered in Ref.~\onlinecite{kavokin05a}. Further, we calculate the \emph{quantum corrections} to polarization conversion and demonstrate that they can either increase of decrease conversion efficiency.

\subsection{Classical effects}\label{sec:cl}

Under the condition Eq. \eqref{generation}, the solution of Eq. \eqref{fk_g} for $f^0(\varphi)$ reads
\begin{equation}\label{f_phi_cl}
f^0(\varphi) = \frac{g\tau}{\mathcal D} \left[ \frac{\tau_0}{\tau_1} + 2\pi \delta(\varphi - \varphi_{\bm k_0}) \right].
\end{equation}
The solution of Eq.~\eqref{sk_g} for $\bm s(\varphi)$ can be written in a straightforward way as well using Eqs. \eqref{s_eq_simple}, \eqref{s_phi1}.
In what follows we concentrate on the important case of excitation
of the pure TE- or TM-state, i.e. $\bm \Omega_{\bm k_0} \parallel
\bm g \parallel x$. Thus, $\overline{\bm s^0} = \bm g u\tau/{\mathcal D}$, and
\begin{eqnarray}
s_x^0(\varphi) &=& \frac{g\tau}{\mathcal D}  \left[2\pi \delta(\varphi) + (1+ \Omega_x^2\tau^2)\nu  u  \right]\label{sx},\\
s_y^0(\varphi) &=& \frac{g\tau}{\mathcal D} \Omega_x\Omega_y\tau^2 \nu u \nonumber,\\
s_z^0(\varphi) &=& -\frac{g\tau}{\mathcal D} \Omega_y\tau\nu u \nonumber,
\end{eqnarray}
where we introduced
\[
u = \left[1+ (\Omega\tau)^2 - \nu (1+ \Omega^2\tau^2/2)\right]^{-1}, \quad \nu = \frac{\tau}{\tau_1}.
\]
\begin{figure}
\centering
\includegraphics[width=\linewidth]{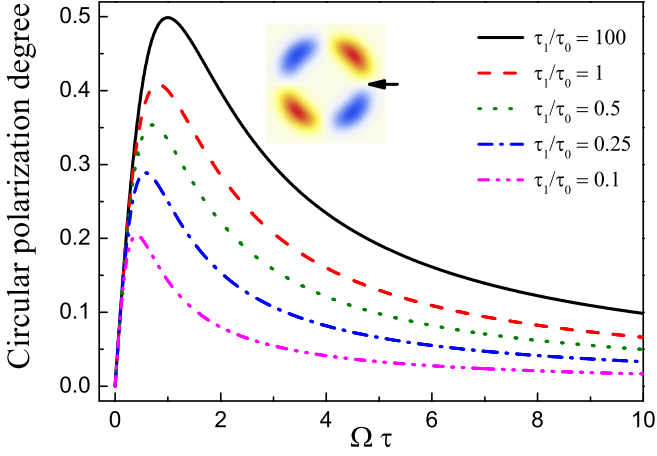}
\caption{(Color online) The absolute value of the circular polarization degree of polaritons as a function of $\Omega\tau$ calculated for the different values of $\tau_0/\tau_1$. The scattering angle is $\pi/4$. An inset shows a contour plot of the angular distribution of circular polarization degree, and arrow denotes the excitation point in the $\bm k$-space.}\label{fig:she}
\end{figure}

The angular distribution of the scattered polaritons is symmetric with respect to the rotation by an angle $\pi$ because the longitudinal-transverse spitting contains even harmonics of the wave vector angle $\varphi_{\bm k}$. Moreover, $s_z$ pseudospin component appears to be proportional to the $y$-component of the splitting, Eq. \eqref{sx}, (see also Ref.~\onlinecite{kavokin05a}), therefore the maxima of circular polarization appear at scattering angles equal to $\pi/4$, $3\pi/4$, $5\pi/4$ or $7\pi/4$. The sign of polarization is opposite in the adjacent maxima.
Figure~\ref{fig:she} shows the circular polarization degree of the polaritons at the scattering by $\pi/4$. The dependence of the polarization degree on $\Omega\tau$ is non-monotonous, its maximum shifts to the lower values of $\Omega\tau$ with decrease of the scattering time~$\tau_1$. The maximum value of the polarization is observed for $\tau_1/\tau_0\to \infty$, i.e. in the regime of a single scattering described in Ref.~\onlinecite{kavokin05a}.

Multiple scattering leads to spin relaxation of polaritons. At $\tau_0 \gg \tau_1$, $(\Omega^2\tau_1)^{-1}$ one can see that the factor $u$ reduces to $2/(\Omega\tau_1)^2$ which is nothing but the ratio of the classical value of the transverse spin relaxation time $\tau_\perp$ and the scattering time $\tau_1$. The circular polarization degree in the scattered state
\[
 \rho_c^0(\varphi)  = \frac{s_z^0(\varphi)}{f^0(\varphi)}= \Omega_y(\varphi) \tau_1 \frac{\tau_\perp}{\tau_0}
\]
is the smaller the shorter spin relaxation time.

\subsection{Quantum effects}\label{sec:quantum}

The quantum corrections to the particle and spin distribution
functions in the lowest order in $W_0$, $\hat W$ can be found
similarly to Sec.~\ref{sec:spinrel}. The distribution functions
are represented as the sum of zeroth order contributions and the
first order corrections, see Eq.~\eqref{decomposition}. Functions
$f^0$ and $\bm s^0$ are given by Eqs. \eqref{f_phi_cl},
\eqref{sx}. Thus, for $f^1(\varphi)$ we have
\begin{equation}\label{f_phi_q}
f^1(\varphi) = \tau W_0 \frac{g\tau}{\mathcal D} \left[2\pi \delta(\varphi - \pi) - 1 \right]
\end{equation}
The first term in the brackets describes the coherent
backscattering, i.e. an increase by interference effects of the number of the particles
scattered into the opposite from the source direction. The second term describes the coherent
scattering by an arbitrary angle. The total number of particles is
conserved, $\int_0^{2\pi} f^1(\varphi) \mathrm d\varphi=0$.

The solution procedure for $\bm s^1(\varphi)$ is analogous to the outlined above. We introduce the auxiliary function
\begin{equation}\label{F1}
\bm F^1(\varphi) = \frac{\overline{\bm s^1}}{\tau_1} + \hat W
\left[\bm s^0 (\varphi - \pi) - \overline{\bm s^0} \right],
\end{equation}
and the kinetic equation for $\bm s^1$ reduces to
Eq.~\eqref{s_eq_simple} with $\bm F^1$ instead of $\bm F$ and $\bm
s^1$ instead of $\bm s$. Thus, $\bm s^1$ is given by
Eq.~\eqref{s_phi1}. One needs to average the solution over
$\varphi$ to obtain self-consistency equations for the components
of $\overline{\bm s^1}$ and then find $\bm s^1(\varphi)$. The
result reads
\begin{eqnarray}
s^1_x(\varphi) &=& \frac{g\tau}{\mathcal D} \biggl[2\pi\tau W_\perp \delta(\varphi-\pi) +  \nonumber\\
& & \frac{u^2}{4} A_0 + \frac{u^2}{4} A_2 \Omega_x^2\tau^2\biggr],\label{s1x}\\
s^1_y(\varphi) &=& \frac{g\tau}{\mathcal D} \frac{u^2}{4} \frac{\Omega_x\Omega_y\tau^2}{2} C_0,\nonumber\\
s^1_z(\varphi) &=& -\frac{g\tau}{\mathcal D} \frac{u^2}{4} \frac{\Omega_y\tau}{2} B_0,\nonumber
\end{eqnarray}
Here the following quantities are introduced:
\[
A_0 = 2\tau W_\parallel  \tau^2(\nu-2)\nu\Omega^2\tau^2 + 4\tau W_\perp (1-\nu)(\nu - 1 - \Omega^2\tau^2),
\]
\[
A_2 = 4\tau W_\parallel (1-\nu)\nu-2\tau W_\perp p,
\]
\[
C_0 = 8\tau W_\parallel (1-\nu)\nu - 4\tau W_\perp p,
\]
\[
B_0 = 8\tau W_\parallel (1-\nu)\nu + 8 \tau W_\perp(1-\nu)(\nu - 1 - \Omega^2\tau^2),
\]
\[
p = \Omega^2\tau^2(2+ \nu^2-4\nu) + 2+ 4\nu^2-6\nu.
\]

Quantum interference leads to the appearance of extra
backscattered polaritons. The total number of backscattered
particles is proportional to $W_0$, Eq.\eqref{f_phi_q}. This
increase is compensated by the coherent scattering by an arbitrary
angle which leads to the decrease of the number of scattered
polaritons in all other directions than exactly backwards.
An increase of the spin splitting at fixed $\tau_1$ and $\tau_0$ leads to the decrease of $W_0$ therefore the number of backscattered particles decreases.

The coherent backscattering is also pronounced in the $x$ pseudospin component. In this case the overall intensity of the backscattering peak is proportional to $W_\perp$.

\begin{figure}
\centering
\includegraphics[width=\linewidth]{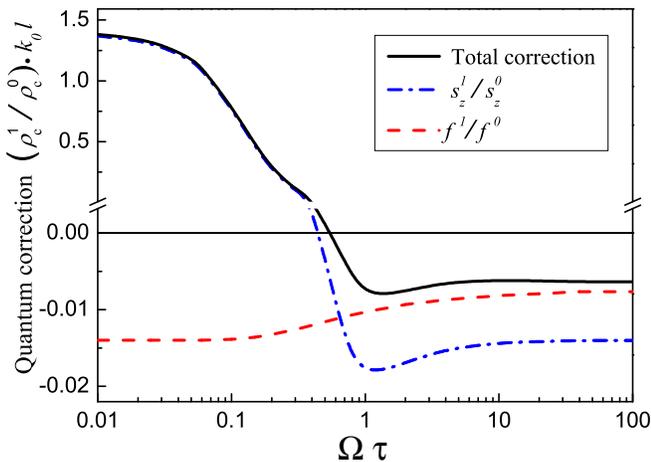}
\caption{(Color online) Quantum corrections to the circular polarization degree
observed under optical spin Hall effect conditions. Solid curve
shows the total correction, dash-dotted one shows first contribution, $s_z^1/s_z^0$, in Eq. \eqref{circ_qnt},
the dashed curve shows second term, $f^1/f^0$,  in Eq.
\eqref{circ_qnt}, $\tau_0/\tau_1 = 100$.}\label{fig:she1}
\end{figure}

The cross-linear polarization and the circular polarization appear proportionally to $\Omega_x\Omega_y$ and $\Omega_y$, therefore both of them vanish for $\varphi=\pi$, i.e. for the detection in the backscattering geometry. However, the coherent scattering by an arbitrary angle leads to the modification of the polarization conversion efficiency. The circular polarization degree can be written as $\rho_c(\varphi) = \rho_c^0(\varphi) + \rho_c^1(\varphi)$, where $\rho_c^0$ is the classical value of the circular polarization degree and the quantum correction $\rho_c^1(\varphi)$ is given by
\begin{equation}\label{circ_qnt}
 \rho_c^1(\varphi) = \rho_c^0(\varphi)\left(\frac{s_z^1}{s_z^0} -\frac{f^1}{f^0}\right).
\end{equation}
The relative value of the polarization conversion efficiency $\rho_c^1/\rho_c^0$ as a function of $\Omega\tau$ is plotted in Fig.~\ref{fig:she1}.

From Eq. \eqref{circ_qnt} it is clear that there are two contributions to the quantum correction to the polarization degree: first one arises due to the modification of spin distribution ($s_z^1$) while second one is determined by the change of the number of particles in a given state.
The latter correction is always positive because the coherent scattering by an arbitrary angle reduces the number of particles in a given state, see Eq.~\eqref{f_phi_q} and dashed curve in Fig.~\ref{fig:she1}. The former one can be either positive or negative depending on the values of $\Omega\tau$ and $\tau_0/\tau_1$, see Fig.~\ref{fig:she1}. For instance, if $\Omega\tau\ll 1$ and $\tau_0\gg \tau_1$ both $W_\parallel$, $W_\perp$ are positive and $B_0>0$. Therefore, quantum corrections in this regime increase the polarization as compared with the classical result. On the contrary, if $\Omega\tau\gg 1$ both $W_\parallel$ and $W_\perp (\nu - 1 - \Omega^2\tau^2) \approx -W_\perp \Omega^2\tau^2$ are negative and $z$-pseudospin component is decreased. Thus, interference of polaritons can either increase of decrease the polarization conversion efficiency.

Physically, it can be interpreted as follows. The efficiency of the polarization conversion is strongly sensitive to the spin relaxation times. The correction to the transverse relaxation time can either be positive or negative depending on the sign of $W_\parallel$ and the value $W_\perp (\Omega\tau)^2$, therefore the $x$-pseudospin component can be preserved better or worse depending on the value of $\Omega\tau$. Therefore generated circular polarization and cross-linear polarization may either increase or decrease as a result of quantum interference.

\section{Conclusions}

To summarize, we have studied in detail the spin-dependent quantum interference and classical multiple scattering effects in dynamics of exciton-polaritons. We have derived the quantum corrections to the collision integral of exciton-polaritons in the leading order in $(k_0l)^{-1}$. The quantum corrections are strongly sensitive to the value of the spin splitting of exciton-polaritons. Contrary to the case of electrons where the strong spin splitting can lead to the anti-localization, the quantum correction to the polariton diffusion coefficient is negative. The quantum correction to spin $z$-component diffusion coefficient changes its sign from negative to positive with the increase of the spin splitting while the correction to the diffusion coefficient of the in-plane spin components is negative. The relaxation of the longitudingal spin component is accelerated by the quantum interference effects and the relaxation rate of the transverse spin components can increase or decrease depending on the spin splitting value. The polarization conversion efficiency in the regime of the optical spin Hall effect can also be larger or smaller than the value predicted by the classical theory depending on the relations between the lifetime of polaritons, their scattering time and the value of the spin splitting.

\acknowledgments{The discussions with A.V. Kavokin, A.N. Poddubny and I.A. Shelykh are gratefully acknowledged. This work was partially supported by RFBR, `Dynasty' Foundation-ICFPM and RSSF.}


\begin{thebibliography}{24}
\expandafter\ifx\csname natexlab\endcsname\relax\def\natexlab#1{#1}\fi
\expandafter\ifx\csname bibnamefont\endcsname\relax
  \def\bibnamefont#1{#1}\fi
\expandafter\ifx\csname bibfnamefont\endcsname\relax
  \def\bibfnamefont#1{#1}\fi
\expandafter\ifx\csname citenamefont\endcsname\relax
  \def\citenamefont#1{#1}\fi
\expandafter\ifx\csname url\endcsname\relax
  \def\url#1{\texttt{#1}}\fi
\expandafter\ifx\csname urlprefix\endcsname\relax\def\urlprefix{URL }\fi
\providecommand{\bibinfo}[2]{#2}
\providecommand{\eprint}[2][]{\url{#2}}

\bibitem[{\citenamefont{Zutic et~al.}(2004)\citenamefont{Zutic, Fabian, and
  Sarma}}]{zutic:323}
\bibinfo{author}{\bibfnamefont{I.}~\bibnamefont{Zutic}},
  \bibinfo{author}{\bibfnamefont{J.}~\bibnamefont{Fabian}}, \bibnamefont{and}
  \bibinfo{author}{\bibfnamefont{S.~D.} \bibnamefont{Sarma}},
  \bibinfo{journal}{Rev. Mod. Phys.} \textbf{\bibinfo{volume}{76}},
  \bibinfo{eid}{323} (\bibinfo{year}{2004}).

\bibitem[{\citenamefont{Kavokin and Malpuech}(2003)}]{kavokin03b}
\bibinfo{author}{\bibfnamefont{A.}~\bibnamefont{Kavokin}} \bibnamefont{and}
  \bibinfo{author}{\bibfnamefont{G.}~\bibnamefont{Malpuech}},
  \emph{\bibinfo{title}{Cavity Polaritons}}, vol.~\bibinfo{volume}{32} of
  \emph{\bibinfo{series}{Thin Films and Nanostructures}}
  (\bibinfo{publisher}{Elsevier}, \bibinfo{year}{2003}).

\bibitem[{\citenamefont{Solnyshkov et~al.}(2007)\citenamefont{Solnyshkov,
  Shelykh, Glazov, Malpuech, Amand, Renucci, Marie, and
  Kavokin}}]{Solnyshkov07}
\bibinfo{author}{\bibfnamefont{D.}~\bibnamefont{Solnyshkov}},
  \bibinfo{author}{\bibfnamefont{I.}~\bibnamefont{Shelykh}},
  \bibinfo{author}{\bibfnamefont{M.}~\bibnamefont{Glazov}},
  \bibinfo{author}{\bibfnamefont{G.}~\bibnamefont{Malpuech}},
  \bibinfo{author}{\bibfnamefont{T.}~\bibnamefont{Amand}},
  \bibinfo{author}{\bibfnamefont{P.}~\bibnamefont{Renucci}},
  \bibinfo{author}{\bibfnamefont{X.}~\bibnamefont{Marie}}, \bibnamefont{and}
  \bibinfo{author}{\bibfnamefont{A.}~\bibnamefont{Kavokin}},
  \bibinfo{journal}{Semiconductors} \textbf{\bibinfo{volume}{41}}, \bibinfo{pages}{1080}
  (\bibinfo{year}{2007}).

\bibitem[{\citenamefont{Panzarini et~al.}(1999)\citenamefont{Panzarini,
  Andreani, Armitage, Baxter, Skolnick, Astratov, Roberts, Kavokin,
  Vladimirova, and M.A.Kaliteevski}}]{panzarini99}
\bibinfo{author}{\bibfnamefont{G.}~\bibnamefont{Panzarini}},
  \bibinfo{author}{\bibfnamefont{L.~C.} \bibnamefont{Andreani}},
  \bibinfo{author}{\bibfnamefont{A.}~\bibnamefont{Armitage}},
  \bibinfo{author}{\bibfnamefont{D.}~\bibnamefont{Baxter}},
  \bibinfo{author}{\bibfnamefont{M.~S.} \bibnamefont{Skolnick}},
  \bibinfo{author}{\bibfnamefont{V.~N.} \bibnamefont{Astratov}},
  \bibinfo{author}{\bibfnamefont{J.~S.} \bibnamefont{Roberts}},
  \bibinfo{author}{\bibfnamefont{A.~V.} \bibnamefont{Kavokin}},
  \bibinfo{author}{\bibfnamefont{M.~R.} \bibnamefont{Vladimirova}},
  \bibnamefont{and} \bibinfo{author}{\bibnamefont{M.A. Kaliteevski}},
  \bibinfo{journal}{Phys. Solid State} \textbf{\bibinfo{volume}{41}},
  \bibinfo{pages}{1223} (\bibinfo{year}{1999}).

\bibitem[{\citenamefont{Kavokin et~al.}(2005)\citenamefont{Kavokin, Malpuech,
  and Glazov}}]{kavokin05a}
\bibinfo{author}{\bibfnamefont{A.}~\bibnamefont{Kavokin}},
  \bibinfo{author}{\bibfnamefont{G.}~\bibnamefont{Malpuech}}, \bibnamefont{and}
  \bibinfo{author}{\bibfnamefont{M.}~\bibnamefont{Glazov}},
  \bibinfo{journal}{Phys. Rev. Lett.} \textbf{\bibinfo{volume}{95}},
  \bibinfo{eid}{136601} (\bibinfo{year}{2005}).

\bibitem[{\citenamefont{Kavokin et~al.}(2004)\citenamefont{Kavokin, Shelykh,
  Kavokin, Malpuech, and Bigenwald}}]{kavokin:017401}
\bibinfo{author}{\bibfnamefont{K.~V.} \bibnamefont{Kavokin}},
  \bibinfo{author}{\bibfnamefont{I.~A.} \bibnamefont{Shelykh}},
  \bibinfo{author}{\bibfnamefont{A.~V.} \bibnamefont{Kavokin}},
  \bibinfo{author}{\bibfnamefont{G.}~\bibnamefont{Malpuech}}, \bibnamefont{and}
  \bibinfo{author}{\bibfnamefont{P.}~\bibnamefont{Bigenwald}},
  \bibinfo{journal}{Phys, Rev. Lett.} \textbf{\bibinfo{volume}{92}},
  \bibinfo{eid}{017401} (\bibinfo{year}{2004}).

\bibitem[{\citenamefont{Mart\'in et~al.}(2002)\citenamefont{Mart\'in, Aichmayr,
  Vi\~na, and Andr\'e}}]{martin}
\bibinfo{author}{\bibfnamefont{M.~D.} \bibnamefont{Mart\'in}},
  \bibinfo{author}{\bibfnamefont{G.}~\bibnamefont{Aichmayr}},
  \bibinfo{author}{\bibfnamefont{L.}~\bibnamefont{Vi\~na}}, \bibnamefont{and}
  \bibinfo{author}{\bibfnamefont{R.}~\bibnamefont{Andr\'e}},
  \bibinfo{journal}{Phys. Rev. Lett.} \textbf{\bibinfo{volume}{89}},
  \bibinfo{pages}{077402} (\bibinfo{year}{2002}).

\bibitem[{\citenamefont{Leyder et~al.}(2007)\citenamefont{Leyder, Romanelli,
  Karr, Giacobino, Liew, Glazov, Kavokin, Malpuech, and Bramati}}]{leyder07}
\bibinfo{author}{\bibfnamefont{C.}~\bibnamefont{Leyder}},
  \bibinfo{author}{\bibfnamefont{M.}~\bibnamefont{Romanelli}},
  \bibinfo{author}{\bibfnamefont{J.~P.} \bibnamefont{Karr}},
  \bibinfo{author}{\bibfnamefont{E.}~\bibnamefont{Giacobino}},
  \bibinfo{author}{\bibfnamefont{T.~C.~H.} \bibnamefont{Liew}},
  \bibinfo{author}{\bibfnamefont{M.~M.} \bibnamefont{Glazov}},
  \bibinfo{author}{\bibfnamefont{A.~V.} \bibnamefont{Kavokin}},
  \bibinfo{author}{\bibfnamefont{G.}~\bibnamefont{Malpuech}}, \bibnamefont{and}
  \bibinfo{author}{\bibfnamefont{A.}~\bibnamefont{Bramati}},
  \bibinfo{journal}{Nature Physics} \textbf{\bibinfo{volume}{3}},
  \bibinfo{pages}{628} (\bibinfo{year}{2007}).

\bibitem[{\citenamefont{Hikami et~al.}(1980)\citenamefont{Hikami, Larkin, and
  Nagaoka}}]{hikami80}
\bibinfo{author}{\bibfnamefont{S.}~\bibnamefont{Hikami}},
  \bibinfo{author}{\bibfnamefont{A.~I.} \bibnamefont{Larkin}},
  \bibnamefont{and} \bibinfo{author}{\bibfnamefont{Y.}~\bibnamefont{Nagaoka}},
  \bibinfo{journal}{Prog. Theor. Phys.} \textbf{\bibinfo{volume}{63}},
  \bibinfo{pages}{707} (\bibinfo{year}{1980}).

\bibitem[{\citenamefont{Altshuler and Aronov}(1985)}]{aa}
\bibinfo{author}{\bibfnamefont{B.L.}~\bibnamefont{Altshuler}} \bibnamefont{and}
  \bibinfo{author}{\bibfnamefont{A.G.}~\bibnamefont{Aronov}},
  in \bibinfo{chapter}{\textit{Electron-electron
  interactions in disordered systems}}, ed. by A.L. Efros and M. Pollak, (North-Holland, Amsterdam, \bibinfo{year}{1985}).

\bibitem[{\citenamefont{Lyubinskiy and Kachorovskii}(2004)}]{lyubinskiy04}
\bibinfo{author}{\bibfnamefont{I.~S.} \bibnamefont{Lyubinskiy}}
  \bibnamefont{and}
  \bibinfo{author}{\bibfnamefont{V.Yu.}~\bibnamefont{Kachorovskii}},
  \bibinfo{journal}{Phys. Rev. B} \textbf{\bibinfo{volume}{70}},
  \bibinfo{pages}{205335} (\bibinfo{year}{2004}).

\bibitem{ivch_pikus} E. L. Ivchenko, G. E. Pikus, B. S. Razbirin, and A. I. Starukhin, Sov. Phys. JETP {\bf 45}, 1172 (1977).

\bibitem[{\citenamefont{Mal\char39{}shukov
  et~al.}(1995)\citenamefont{Mal\char39{}shukov, Chao, and
  Willander}}]{PhysRevB.52.5233}
\bibinfo{author}{\bibfnamefont{A.~G.} \bibnamefont{Mal\char39{}shukov}},
  \bibinfo{author}{\bibfnamefont{K.~A.} \bibnamefont{Chao}}, \bibnamefont{and}
  \bibinfo{author}{\bibfnamefont{M.}~\bibnamefont{Willander}},
  \bibinfo{journal}{Phys. Rev. B} \textbf{\bibinfo{volume}{52}},
  \bibinfo{pages}{5233} (\bibinfo{year}{1995}).

\bibitem[{\citenamefont{Savona et~al.}(2000)\citenamefont{Savona, Runge, and
  Zimmermann}}]{Savona00}
\bibinfo{author}{\bibfnamefont{V.}~\bibnamefont{Savona}},
  \bibinfo{author}{\bibfnamefont{E.}~\bibnamefont{Runge}}, \bibnamefont{and}
  \bibinfo{author}{\bibfnamefont{R.}~\bibnamefont{Zimmermann}},
  \bibinfo{journal}{Phys. Rev. B} \textbf{\bibinfo{volume}{62}},
  \bibinfo{pages}{R4805} (\bibinfo{year}{2000}).

\bibitem[{\citenamefont{Gurioli et~al.}(2005)\citenamefont{Gurioli, Bogani,
  Cavigli, Gibbs, Khitrova, and Wiersma}}]{gurioli:183901}
\bibinfo{author}{\bibfnamefont{M.}~\bibnamefont{Gurioli}},
  \bibinfo{author}{\bibfnamefont{F.}~\bibnamefont{Bogani}},
  \bibinfo{author}{\bibfnamefont{L.}~\bibnamefont{Cavigli}},
  \bibinfo{author}{\bibfnamefont{H.}~\bibnamefont{Gibbs}},
  \bibinfo{author}{\bibfnamefont{G.}~\bibnamefont{Khitrova}}, \bibnamefont{and}
  \bibinfo{author}{\bibfnamefont{D.~S.} \bibnamefont{Wiersma}},
  \bibinfo{journal}{Phys. Rev. Lett.} \textbf{\bibinfo{volume}{94}},
  \bibinfo{eid}{183901} (\bibinfo{year}{2005}).

\bibitem[{\citenamefont{Maialle et~al.}(1993)\citenamefont{Maialle,
  de~Andrada~e Silva, and Sham}}]{maialle93}
\bibinfo{author}{\bibfnamefont{M.}~\bibnamefont{Maialle}},
  \bibinfo{author}{\bibfnamefont{E.}~\bibnamefont{de~Andrada~e Silva}},
  \bibnamefont{and} \bibinfo{author}{\bibfnamefont{L.}~\bibnamefont{Sham}},
  \bibinfo{journal}{Phys. Rev. B} \textbf{\bibinfo{volume}{47}},
  \bibinfo{pages}{15776} (\bibinfo{year}{1993}).

\bibitem[{\citenamefont{Dmitriev et~al.}(1997)\citenamefont{Dmitriev,
  Kachorovskii, and Gornyi}}]{dmitriev97}
\bibinfo{author}{\bibfnamefont{A.~P.} \bibnamefont{Dmitriev}},
  \bibinfo{author}{\bibfnamefont{V.~Y.} \bibnamefont{Kachorovskii}},
  \bibnamefont{and} \bibinfo{author}{\bibfnamefont{I.~V.}
  \bibnamefont{Gornyi}}, \bibinfo{journal}{Phys. Rev. B}
  \textbf{\bibinfo{volume}{56}}, \bibinfo{pages}{9910} (\bibinfo{year}{1997}).

\bibitem[{\citenamefont{Savona}(2007)}]{savona07}
\bibinfo{author}{\bibfnamefont{V.}~\bibnamefont{Savona}}, \bibinfo{journal}{J.
  Phys.: Condens. Matter} \textbf{\bibinfo{volume}{19}},
  \bibinfo{pages}{295208} (\bibinfo{year}{2007}).

\bibitem[{\citenamefont{Iordanskii et~al.}(1994)\citenamefont{Iordanskii,
  Lyanda-Geller, and Pikus}}]{iordanskii94}
\bibinfo{author}{\bibfnamefont{S.V.}~\bibnamefont{Iordanskii}},
  \bibinfo{author}{\bibfnamefont{Y.B.}~\bibnamefont{Lyanda-Geller}},
  \bibnamefont{and} \bibinfo{author}{\bibfnamefont{G.E.}~\bibnamefont{Pikus}},
  \bibinfo{journal}{JETP Letters}
  \textbf{\bibinfo{volume}{60}}, \bibinfo{pages}{199} (\bibinfo{year}{1994}).

\bibitem{golub_wl_05} L.E. Golub, Phys. Rev. B \textbf{71}, 235310 (2005).

\bibitem[{\citenamefont{Glazov and Golub}(2006)}]{glazov_wl_06}
\bibinfo{author}{\bibfnamefont{M.M.}~\bibnamefont{Glazov}} \bibnamefont{and}
  \bibinfo{author}{\bibfnamefont{L.E.}~\bibnamefont{Golub}},
  \bibinfo{journal}{Semiconductors} \textbf{\bibinfo{volume}{40}},
  \bibinfo{pages}{1209} (\bibinfo{year}{2006}).


\bibitem[{\citenamefont{Arseev and Dzyubenko}(1998)}]{arseev98}
\bibinfo{author}{\bibfnamefont{P.I.}~\bibnamefont{Arseev}} \bibnamefont{and}
  \bibinfo{author}{\bibfnamefont{A.B.}~\bibnamefont{Dzyubenko}},
  \bibinfo{journal}{JETP} \textbf{\bibinfo{volume}{87}}, \bibinfo{pages}{200}
  (\bibinfo{year}{1998}).

\end{thebibliography}

\end{document}